\documentclass[showpacs,preprintnumbers,pre]{revtex4}
\usepackage{bm}
\usepackage{epsf}
\begin{document}

\title{Scaling Analysis of Surfactant Templated Polyacrylamide Gel Surfaces}
\author{
Mukundan Chakrapani$^1$, S.~J. Mitchell$^{1,2}$, D.~H.
Van~Winkle$^1$, and P.~A. Rikvold$^{1,2}$
}
\affiliation{
$^1$Center for Materials Research and Technology 
and Department of Physics, 
Florida State University, Tallahassee, Florida 32306-4351\\
$^2$School of Computational Science and Information Technology,
Florida State University, Tallahassee, Florida 32306-4120
}

\date{\today}

\begin{abstract}
Surfaces of  surfactant-templated polyacrylamide hydrogels were imaged 
by atomic force microscopy (AFM), and the surface morphology was studied by 
numerical scaling analysis. The templated gels were formed by polymerizing 
acrylamide plus a cross-linker in the presence of surfactants, which were 
then removed by soaking in distilled water.
Gels formed in the presence of over $20\%$ surfactant (by weight) formed clear,
but became opaque upon removal of the surfactants.
Untemplated gels formed and remained clear. 
The surface morphology of the gels was studied by several one- and 
two-dimensional numerical scaling methods. 
The surfaces were found to be self-affine on short length scales, 
with a roughness (Hurst) exponent in the range 0.85 to 1, 
crossing over to a constant root-mean-square surface width on long scales. 
Both the crossover length between these two regimes and the saturation value 
of the surface width increased significantly with increasing surfactant 
concentration, coincident with the increase in opacity. We propose that the 
changes in the surface morphology are due to a percolation transition in the 
system of voids formed upon removal of the surfactants from the bulk. 
\end{abstract}
\maketitle

\section{Introduction}
\label{sec:intro}

Polyacrylamide (PAAm) gels are used extensively for separations of
biological macro-molecules. 
Because of their intrinsically broad distribution of pore sizes
they are very useful for a wide range of separations for protein 
isolation and sequencing of single-stranded DNA~\cite{chiari95}. 
In particular, protein separations achieved by electrophoresis on
gels templated with surfactant micelles have recently been
reported~\cite{rill96}, and the possibility of improvements of the
separation properties, when pores of a particular size are
templated within the gels, have been proposed~\cite{rill98}.
Extensive work has also been done on the possibility of using
such templated gel matrices in gel permeation chromatography
(GPC) \cite{patterson00}.

Since its introduction in 1986~\cite{binnig86}, atomic force
microscopy (AFM) has been widely used for recording images of many surfaces
with resolution at small length scales. 
Recent developments in AFM
technology have enabled its use, not only in high-resolution
profiling of surface morphology and nanostructure, but also in
determining local mechanical properties and in surface compositional
mapping of heterogeneous samples. These applications have been
extensively discussed by Magonov {\em et al.}~\cite{magonov97,magonov98}.
Despite such developments, there have been only a few reports in
the literature on the direct observation of polymer gel surfaces
by AFM~\cite{suzuki96,suzuki97,maaloum98}. Here, an AFM study of
the surface morphologies of templated PAAm hydrogels and
their scaling properties are presented.

\begin{figure}
{\epsfxsize=4in \epsfbox{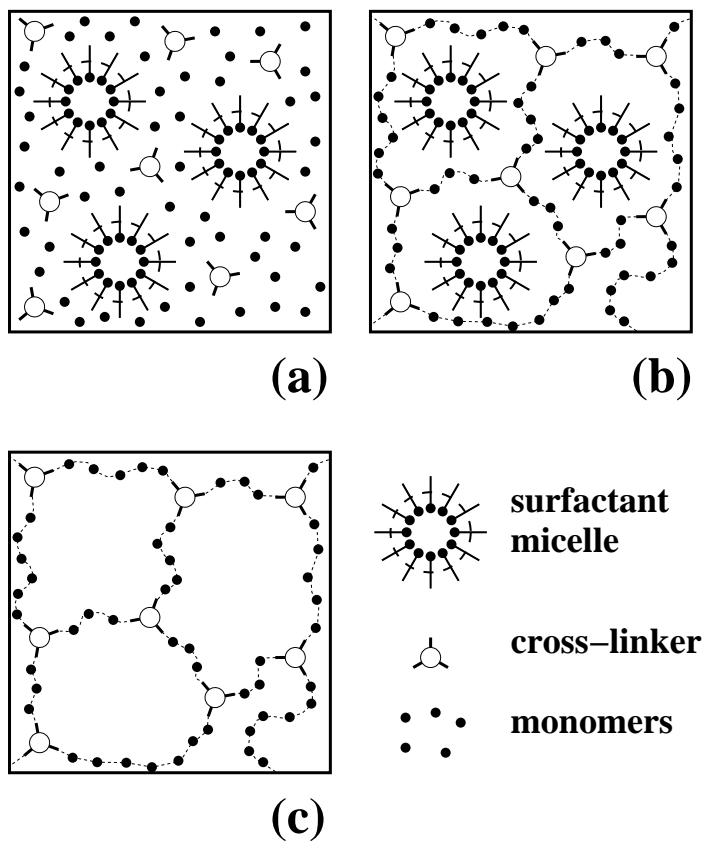}}
\caption{Schematic representation of templated-gel synthesis. 
(a) In step one, monomers, cross-linkers, and surfactant micelles are mixed in solution.
(b) In step two, polymerization occurs around the unreactive surfactants.
(c) In the final step, the surfactants are removed by soaking in distilled water,
leaving behind templated voids.
The thin dashed lines represent chemical bonds between monomers and cross-linkers.
The remaining symbols are explained in the figure legend.
}
\label{fig:template}
\end{figure}

We use micelles of tetradecyltrimethyl ammonium bromide (TTAB)
surfactant to template pores in PAAm.  Aqueous
solutions of surfactant, acrylamide, and N',N'-methylene
bisacrylamide (cross-linker) are mixed at various concentrations
that allow the surfactant molecules to self-assemble into
micelles. The solutions are then polymerized, after which the
surfactants are removed by soaking in distilled water to produce templated
hydrogels. A schematic representation of the templating process is 
shown in Fig.~\ref{fig:template}.
In previous applications of TTAB templated PAAm gels 
for GPC~\cite{patterson00}, mean pore sizes, exclusion
limits, and mechanical integrity were determined. It was shown
that templated gels yield improved resolution for the separation
of globular proteins from 10,000 to 150,000 Da  molecular weight.
Electrophoretic separations achieved on templated PAAm were also
shown to be consistent with the pore-size distributions observed
by GPC analysis. Templated PAAm gel matrices have the
potential to improve separation efficiency over a wide range of
globular protein sizes.

In this paper we present a quantitative study of the structural changes due
to the templating process, using scaling analysis of AFM images of
the surfaces of templated and untemplated gels. A wide variety of
image-analysis tools are available with the Digital Instruments
AFM software~\cite{crmdi3000}. However, these are not well
suited for scaling analysis of soft surfaces. Hence, we
investigated the surface structure over a wide range of TTAB
concentrations using our own, custom-designed surface-analysis
software. This scaling analysis appears to be a robust tool to
analyze and characterize surfaces imaged by AFM.

The rest of this paper is organized as follows. In Sec.~\ref{sec:exp} we
describe the experimental procedure, apparatus, and results. In
Sec.~\ref{sec:meth} we discuss the scaling-analysis methods. This
section is subdivided into two-dimensional (Sec.~\ref{sec:2d})
and one-dimensional (Sec.~\ref{sec:1d}) methods. In
Sec.~\ref{sec:results} we discuss the results of the scaling
analysis of the AFM images, and in Sec.~\ref{sec:conc} we present
our conclusions.

\section{Experimental}
\label{sec:exp}

PAAm gels were formed at room temperature 
by free-radical polymerization with chemical initiators
from a mixture of monomer, cross-linker, buffer, and surfactant. 
The gels were cast
between two glass slides separated by $1.5$~mm spacers. 
The glass slides were ordinary precleaned microscope slides (Fisher Scientific). 
They were cleaned with detergent and rinsed in distilled water 
and air dried before use.
AFM images (not shown here) of these slides revealed an 
root-mean-square (rms) roughness of less than a nanometer over
several square micrometers.
Acrylamide (Electrophoresis Grade, Fisher Scientific) solutions were 
polymerized in the presence of
a TBE buffer (45~mM tris-borate, 1~mM EDTA) at pH~$8.3$. The
chemical initiators were ammonium persulfate (Ammonium Peroxydisulfate, 
Certified A.C.S., Fisher Scientific) and N, N, N',
N'-tetramethylethylenediamine (TEMED, Fisher Scientific). The
concentration of acrylamide was kept constant at $40\%$ by weight,
and that of the cross-linker, N',N'-methylene
bisacrylamide (Bis-acrylamide, Electrophoresis Grade, 
Fisher Scientific)
was held at $7\%$ of the acrylamide concentration in all samples. 
Various amounts of the surfactant, TTAB (Sigma-Aldrich), were added 
to form the templated gels.
The amount of surfactant added was so that the final surfactant
concentration ranged from $0\%$ to $40\%$ by weight
in steps of $10 \%$. The presence of
surfactant micelles did not prevent the formation of
PAAm gels. The rates of polymerization were, however,
altered by the presence of surfactants.  Higher concentration of
surfactants in the pre-gel solution reduced the rate of acrylamide
polymerization. All pre-gel solutions were clear, homogeneous
solutions with viscosities dependent on the amount of surfactant.
The resulting templated gels were all optically clear before
surfactant removal.

After polymerization, the gels were peeled off the glass plates and
soaked in distilled water to remove the surfactants by diffusion. 
More than $98\%$ of the surfactants diffused
out of the gel upon soaking for about two
days. Several methods, including Raman spectroscopy have been used to 
quantify the removal of the surfactant templates by free
diffusion into water~\cite{patterson00}. During the
surfactant-removal process, gels with surfactant concentrations of
more than $20\%$ transformed into white homogeneous materials.
This suggests a structural change in the templated gels, involving
a length scale comparable to the wavelength of visible light. This
further indicates that the structural change is associated with
kinetic restructuring of the gel during and after the removal of micelles.

\begin{figure}
{\epsfxsize=4in \epsfbox{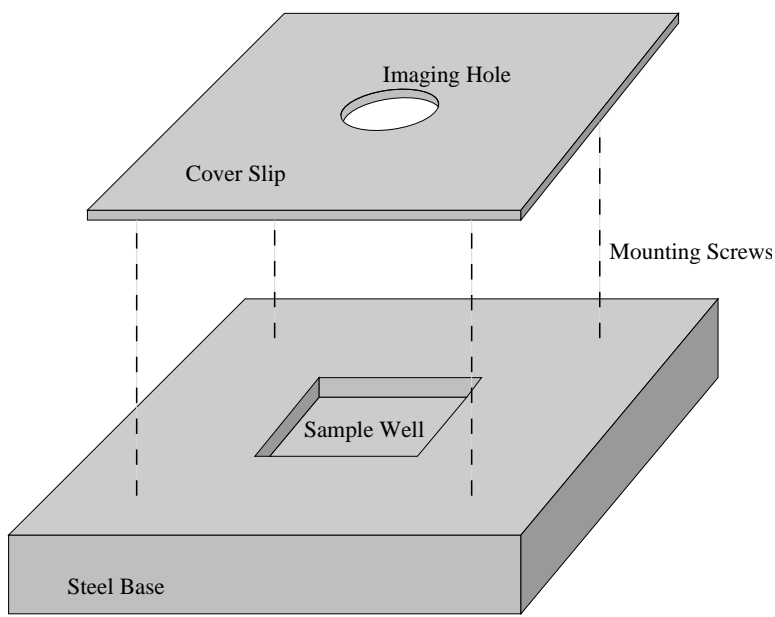}}
\caption{ Schematic representation of the sample holder designed
for AFM imaging of hydrogels. }
\label{fig:holder}
\end{figure}

The gel surfaces were imaged using AFM (Digital Instruments, D3000) with a  silicon
nitride probe (Digital Instruments) of spring constant 0.32~N/m. Scans
were performed in contact mode under water (HPLC grade). 
The set-point voltage was kept at the lowest possible value to
minimize the risk of damage to the gel. The inherent soft nature
of the gels under investigation can cause difficulties in imaging
and affect the reproducibility of the images. In order to overcome this
difficulty, we devised a sample holder which helped provide the
required stability and hence improved the reproducibility of the images. This
sample holder consists of a stainless-steel block with a well in
the center, into which the sample is placed. The sample is then
covered using a thin steel cover slip with a hole in the center
for imaging. The cover slip is held in place with mounting screws 
(see Fig.~\ref{fig:holder}).

\begin{figure}
{\epsfxsize=4in \epsfbox{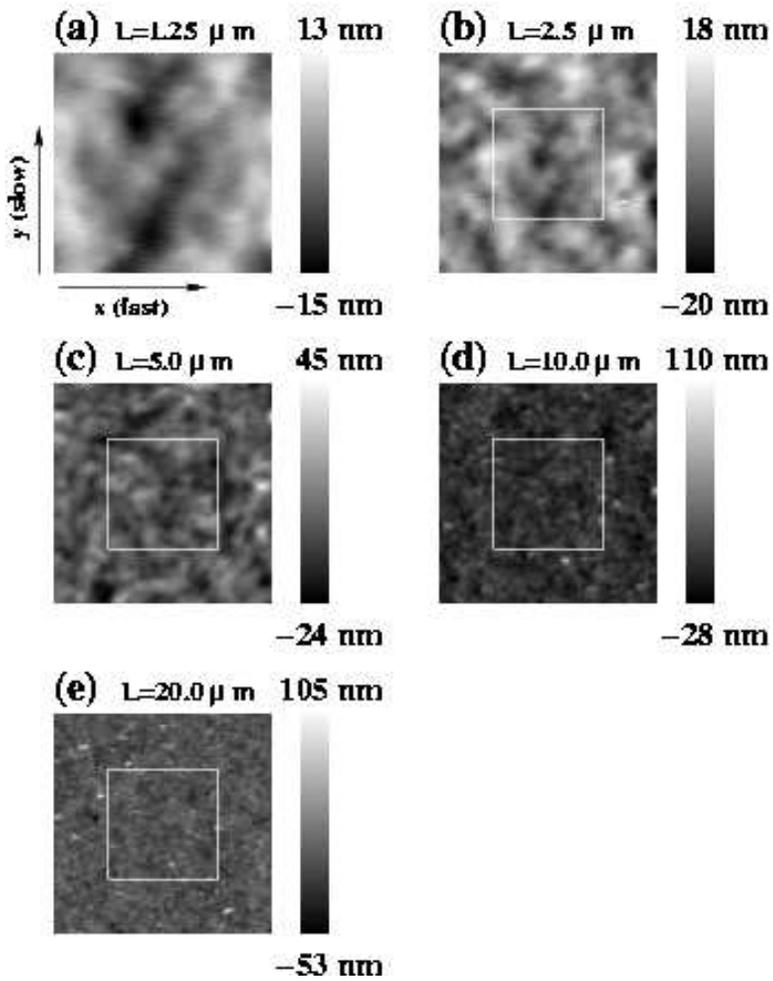}} 
\caption{ 
AFM images of a
set of successively zoomed-out images centered at the same
location on a $20\%$ templated gel surface. Parts (a) through (e)
represent $L \times L$ scan sizes with $L=$ 1.25, 2.5, 5.0, 10.0,
and 20.0~$\mu \rm{m}$, respectively. The gray scale represents
the surface height, with white (black) corresponding to the
maximum (minimum) height. The lateral drift of the sample was
very small. Note the increase in height range with increasing
$L$. The white squares indicate the next smaller scan areas.
} 
\label{fig:zoom}
\end{figure}

\begin{figure}
{\epsfxsize=4in \epsfbox{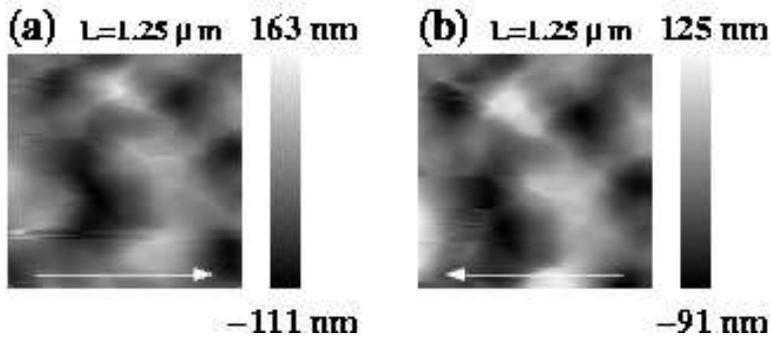}} 
\caption{ 
AFM images of
a $40\%$ templated gel surface at $L=1.25~\mu \rm{m}$ scan size,
(a) forward scan direction, and (b) reverse scan direction. In
each image white (black) corresponds to the maximum (minimum)
height. The distortion in the direction of the tip motion
(indicated by white arrows) is readily observed.
}
\label{fig:scan-rescan}
\end{figure}

After the surfactants were removed by soaking, three different
locations on each gel surface were selected for imaging, and a set
of centered, zoomed images were obtained for each location. The
image sets were created by scanning the surface from small scan
sizes up to larger sizes, in sequence. The images were all $L \times L$ squares
with sides $L$=1.25, 2.5, 5.0, 10.0, and 20.0~$\mu$m,
respectively. This process yielded a set of successively
zoomed-out images of the same location.The lateral drift of the
sample was very small. Figure~\ref{fig:zoom} shows one such zoomed
set, and the excellent reproducibility of the images justifies the
use of the cover slip holder. To further test the reproducibility,
drift, and surface distortion, two images were obtained for
each scan size, first in the forward scanning direction, and next
in the reverse scanning direction, before changing to the next
scan size (see Fig.~\ref{fig:scan-rescan}). Some distortion along the
direction of the tip motion is noticeable. All images have $512
\times 512$ pixel resolution, regardless of scan size, and all
imaging was performed at room temperature.

This telescoping scanning procedure produced a set of ten images
for each location and a total of thirty images for each gel. With
this procedure, five different gels were examined with surfactant
concentrations of 0, 10, 20, 30, and 40$\%$, yielding a
total set of 150 images (75 forward scans and 75 reverse scans).
Many of the images were too noisy for analysis, and all 150 images
were examined for quality in a blind procedure in which two of the
authors, who had not themselves acquired the data, were separately
shown all 150 images. They indicated whether each image should be
retained for analysis. If an image was indicated as too noisy by
either individual, that image was excluded from analysis. This
procedure yielded a total of 106 images for analysis, as detailed in 
Table~\ref{tab:sets}.

The gel surfaces were also imaged using tapping mode under water. 
A single set of ten images as described above was obtained for gels with 
surfactant concentrations of $0\%$ and $40\%$.
The surface morphologies were statistically indistinguishable from those 
obtained using contact mode. 
Since tapping mode is rather time consuming and did not provide any
apparent advantage over contact mode, all the data presented here were 
collected in contact mode.

\begin{figure}
{\epsfxsize=4in \epsfbox{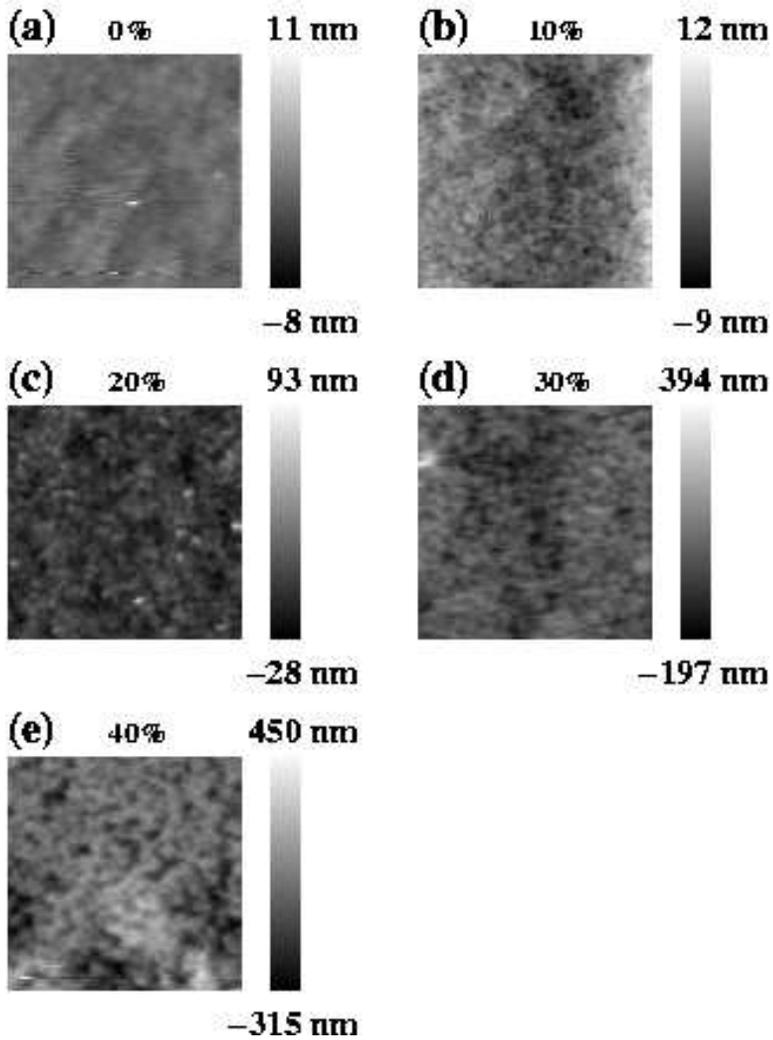}} 
\caption{ 
AFM images of
gel surfaces with surfactant concentrations of (a)~$0\%$,
(b)~$10\%$, (c)~$20\%$, (d)~$30\%$, and (e)~$40\%$. The scan size
is $L=10.0~\mu \rm{m}$ with $512 \times 512$ pixel resolution. In
each image white (black) corresponds to the maximum (minimum)
height. The height range depends strongly on the surfactant
concentration, as is also shown in Figs.~\protect\ref{fig:3dimages10} 
and~\protect\ref{fig:scanline}.
} 
\label{fig:images10}
\end{figure}

\begin{figure}
{\epsfxsize=4in \epsfbox{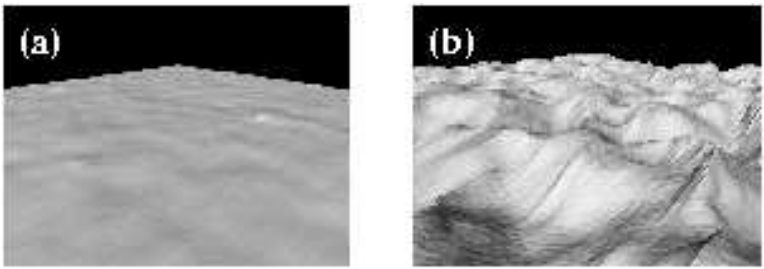}} 
\caption{ 
Three-dimensional views
of AFM images of gel surfaces with $L=10~\mu \rm{m}$ and
surfactant concentrations of (a)~$20\%$, and (b)~$40\%$
(corresponding to Fig.~\ref{fig:images10} (c) and (e),
respectively). The perspective shown is that of an observer
sitting at the center of the image at a height of $1~\mu \rm{m}$
above the average height of the surface and looking down at a
corner. The light source is behind the observer and to the left.
}
\label{fig:3dimages10}
\end{figure}

\begin{figure}
{\epsfxsize=4in \epsfbox{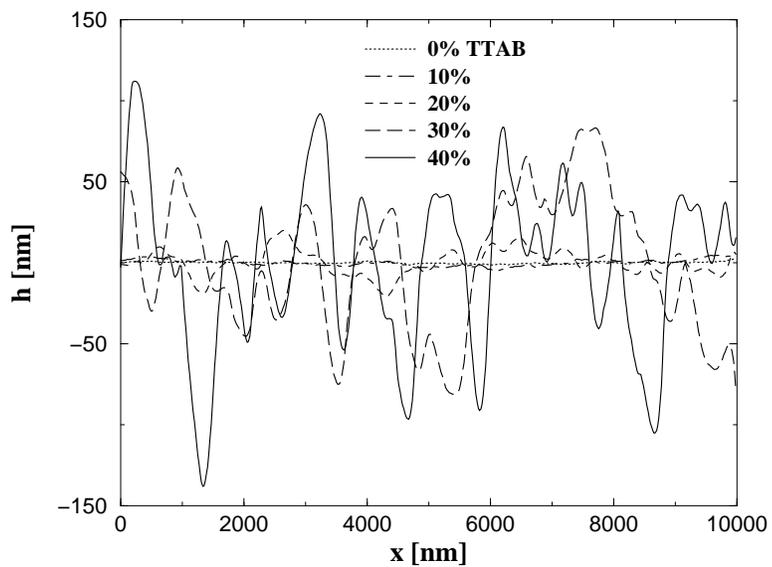}} 
\caption{ 
Plots of typical scan lines for $L=10.0~\mu \rm{m}$ gel surfaces (from
Fig.~\protect\ref{fig:images10}) of various surfactant concentrations.
Note the strong dependence of the height range on the surfactant
concentration.
} 
\label{fig:scanline}
\end{figure}

Figure~\ref{fig:images10} shows examples of AFM images ($L=10.0~\mu
\rm{m}$) of gel surfaces with each of the five different template
concentrations. The gels exhibit marked statistical differences,
most notably that the height range of the surface depends on both
the surfactant concentration and the length scale of the scan.
This is clearly seen in the three-dimensional visualizations of the
gel surface, shown in Fig.~\ref{fig:3dimages10}. Typical scan
lines for gel surfaces at various surfactant concentrations are
shown in Fig.~\ref{fig:scanline}. On the 10~$\mu$m length scale,
the untemplated gels have height ranges on the order of a few nanometers,
while the 40$\%$ templated gels have height ranges on the order of
a few hundred nanometers. AFM images and Raman spectroscopy of glass plates
before and after polymerization and gel removal indicated that there 
was no polymer left behind on the glass. 
This suggests that the gel surfaces imaged by AFM are relaxed, free surfaces,
rather than fracture surfaces. 

The surface analysis routines available with the
D3000 only quantify changes in the overall surface width 
(often called ``the surface roughness''). 
In contrast, our statistical analysis shows self-affine
scaling behavior at small length scales, which crosses over to a
length-scale independent behavior at large length scales, as
discussed in Sec~\ref{sec:results}. As we are interested in the
behavior of the surface on a wide range of length scales, no
image filtering or processing was performed, with the exception of
flattening (described in Sec.~\ref{sec:2d}). It is our opinion 
that scaling analysis of filtered images could be highly suspect
since the essence of most image processing techniques is to alter
the data differently on different length scales. All analysis was
performed with our own code, written in C, as discussed in
Sec.~\ref{sec:meth}.

\section{Analysis Methods}
\label{sec:meth}

We here present a well-defined procedure to characterize surfaces,
based on the statistical analysis methods described in
Refs.~\cite{barabasi95,yang93}. One common measure of surface
morphology is the rms height of the surface,
which is sometimes referred to as the ``rms roughness'' (referred
to here as the rms surface width),
\begin{equation}
w=\sqrt{\frac{\sum_{i=1}^{M^2}{h_i^2}}{M^2}
-\left[\frac{\sum_{i=1}^{M^2}h_i}{{M^2}}\right]^2 }, \label{eq:width}
\end{equation}
where $h_i$ is the $i$th value of the height in an image
consisting of $M \times M$ pixels. (For all of our images,
$M=512$.) However, such a drastic simplification of the
surface structure to only one single number ignores nearly all of
the morphological information contained in the image. 
Reporting the surface width without specifying further details can be meaningless.
The value of the width is highly dependent, not only on the image processing
methods used~\cite{fang97}, but also on the length scale of
observation~\cite{iwasaki93}. We therefore present a comprehensive
scaling analysis of the gel surfaces which goes beyond the simple
single $w$ description.

The surface width defined by Eq.~(\ref{eq:width}) for a $L \times L~(\mu{\rm m}^{2})$ 
image of $M \times M$
pixels varies with $L$. 
For a variety of different surfaces the width is found to exhibit different length
dependences with two limits:
\begin{equation}
w \propto
\left\{
\begin{array}{ll}
L^{\alpha} & L \ll l_{\times} \\
w_{\rm sat} & L \gg l_{\times}
\end{array}
\right. \; ,
\label{eq:behavior}
\end{equation}
where the exponent $\alpha$ is the Hurst exponent (in this study
equal to the roughness exponent~\cite{meakin98}), $w_{\rm sat}$ is
a constant saturated surface width, and $l_{\times}$ is a
cross-over length scale between the two behaviors. In general,
$\alpha$, $l_{\times}$, and $w_{\rm sat}$ depend on the details of the
surface, such as the surfactant concentration in the present study.

The power law in Eq.~(\ref{eq:behavior}) is a consequence of
self-affine scaling behavior, in which the surface is statistically invariant
under the transformation~\cite{barabasi95}
\begin{equation}
\begin{array}{ccc}
x\rightarrow b x, & y\rightarrow b y, & h\rightarrow b^{\alpha} h\;,
\end{array}
\label{eq:trans}
\end{equation}
where $x$ and $y$ are distances in the two image-scan directions,
$h$ is the height, and $b$ is a dimensionless re-scaling factor.
Self-affine behavior is different from self-similar behavior,
which involves invariance under the isotropic scale-change
transformation~\cite{barabasi95} $x\rightarrow bx$, $y\rightarrow
by$, and $h\rightarrow bh$. This distinction between self-affine
and self-similar structures is often overlooked, and the term
``fractal'' is used somewhat freely for both in the literature.

If the surface were ideally self-affine, then the power-law
behavior in Eq.~(\ref{eq:behavior}) would continue on all length
scales, and there would be neither a saturation width, $w_{\rm
sat}$, nor any identifiable length scale such as $l_{\times}$ or a
typical feature size. 
For a surface whose heights are described by an ideal random walk,
$\alpha=1/2$ and $l_{\times}\rightarrow \infty$. 
For most surfaces, measured values of $\alpha$ are
between $1/2$ and $1$~\cite{maloy92, bouchaud93, delaplace99},
even for such different surfaces as brittle rocks
($\alpha \approx 0.78$)~\cite{brown85} and CuCl islands on CaF$_{2}$
($\alpha \approx 0.84$)~\cite{tong94}. As discussed in
Sec.~\ref{sec:results}, our estimates of $\alpha$ for the
different gel surfaces all fall within the range $[0.5,1.0]$, and
$w_{\rm sat}$ and $l_{\times}$ vary strongly with the 
surfactant concentration.

In the following subsections, we discuss several different methods
used to examine the scaling behavior of the gel surfaces. Both
two-dimensional analysis methods that analyze the image as a
whole, and one-dimensional methods that analyze the image one
scan line at a time were examined.

\subsection{Two-Dimensional Methods}
\label{sec:2d}

The AFM records the vertical surface height, $h(x,y)$, as a
function of the fast ($x$) and slow ($y$) scanning directions. As
already discussed, $L$ dependent surface widths can be obtained from $L \times
L~(\mu {\rm m}^{2})$ images. After imaging at a sufficiently large
number of scan sizes (different $L$'s), $\alpha$, $w_{\rm sat}$,
and $l_{\times}$ can be obtained. However, the number of images
required for such an analysis is large. By noting that each image
contains information about length scales from $L/M$ (where $M$ is 
the number of pixels) up to $L$, one
can perform analyses over a wide range of length scales from a
single AFM image. Several variations of two-dimensional analysis
methods were studied. We discuss a spectral method based
on the Fourier transform of the image~\cite{yang93}, a box-counting method from
subsections of images, and widths calculated from the same subsections of the
image~\cite{barabasi95}. 
All of the images were flattened using the AFM software~\cite{crmdi3000} and 
not processed any further. 
The flattening procedure fits a straight line to each scan line by the method of
least squares, which is then subtracted from the data.
The fitting and subtracting procedure was applied, independently,
to each scan line in the image. After flattening, the average
height and slope in the $x$-direction (the fast scan direction) are zero.

\begin{figure}
{\epsfxsize=4in \epsfbox{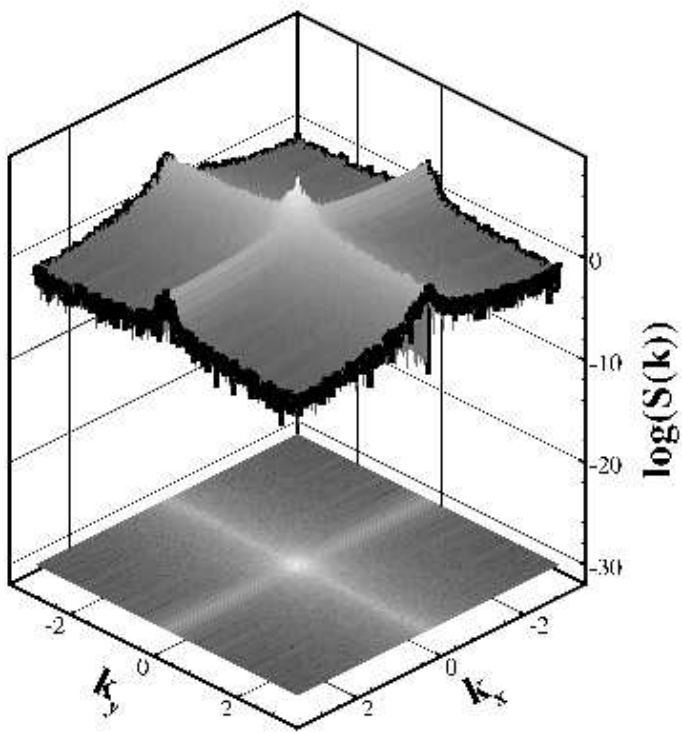}} 
\caption{ 
Structure factor $S(k_x, k_y)$
of a $40\%$ templated gel on a logarithmic gray scale. The structure factor is
given in arbitrary units. The anisotropy in the AFM images, which
appears as large ridges along the $k_x=0$ and $k_y=0$ axes, makes
circular averaging unreliable. 
} 
\label{fig:struct}

\end{figure}

For the spectral methods, we computed the structure factor,
$S(\vec{k}) \propto |{\mathcal F}_{\vec k}(h(x,y))|^2$, which is
the absolute square magnitude of the complex Fourier transform of
the surface height. Here, $\vec{k}$ is the wave vector in the $x,y$
plane. For large $k$, the structure
factor should scale as $S(k) \propto k^{-(2+2 \alpha)}$
\cite{yang93}, a generalized Porod's Law \cite{porod83}. 
The presence of anisotropy in the AFM images produces ridges along the $k_x$
and $k_y$ directions as shown in Fig.~\ref{fig:struct}.
This made circular averaging unreliable, and hence the results from spectral 
methods were suspicious.
The presence of the ridges seems to be a general feature of
AFM imaging, as has been observed for hard materials as well~\cite{fang97}.
Removal of these ridges (which have a non-zero width) by a technique such as 
that suggested in Ref.~\cite{fang97} would require {\em
a priori} knowledge of their functional form. 

For both the box-counting and width methods, the $L \times L~(\mu
{\rm m}^{2})$ image of $M \times M$ height values (pixels) was
recursively subdivided into non-overlapping images of $l \times
l~(\mu {\rm m}^{2})$ with $(M/n) \times (M/n)$ pixels, where
$l=L/n$ $\mu$m and $n\leq{M/2}$ is an integer denoting the level
of subdivision. For each value of $l$, $(L/l)^2$ different
sub-images were obtained.
Non overlapping sub-images were used to ensure statistical independence.

A box-counting method was tried following Ref.~\cite{barabasi95}. 
The minimum number of
cubic boxes of side $l$ needed to span the height range of a
sub-image of $l \times l~\mu {\rm m}^{2}$ was calculated. 
However, the height range was nearly always less than $l$, yielding 
the total number of boxes to be one, regardless of the length scale.
This would seem to imply $\alpha = 1$. 
Thus the box-counting method is entirely insensitive to the scaling 
behavior of the gel surfaces.

The rms width for the $j$-th sub-image of size $l \times l$,
\begin{equation}
w_{l j}=
\sqrt{\frac{\sum_{i=1}^{(M/n)^2}{h_i^2}}{(M/n)^2}
-\left[\frac{\sum_{i=1}^{(M/n)^2}h_i}{{(M/n)^2}}\right]^2 }\;,
\label{eq:widthl}
\end{equation}
and its average, $\langle w_l \rangle$, over all sub-images of
size $l$ were calculated. Typical results are
shown in Fig.~\ref{fig:2dwidth}. 
For a discussion of the effects of changing the order of the 
averaging and the square root, see Ref.~\cite{BUEN02}. 

Among the several two-dimensional analysis methods used to 
study the scaling properties of the gel surfaces,
the calculation of $\langle w_l\rangle$ was found to provide 
the most reliable results.
The other methods have inherent disadvantages as discussed 
in the preceding paragraphs. 
As we shall explore in the next subsection, the scanning 
process introduces noise between the
scan lines, and for a more accurate analysis, one-dimensional
methods should be used.

\subsection{One-Dimensional Methods}
\label{sec:1d}

If there is noise in the measurement process we would expect to
measure $h_{\rm measured}(x,y)=h(x,y)+h_{\rm noise}$, where the
noise adds an effective offset.
If we further assume that $h_{\rm noise}$ is a random walk in
time, noise correlations in time would appear as
correlations in space. Furthermore, if the noise varies
sufficiently fast compared to the scanning velocity, $h_{\rm
noise}$ at the beginning of each scan line will be completely uncorrelated. 
Thus for a perfectly flat surface with $h(x,y)=0$,
each scan line would appear to be a random walk with $\alpha=1/2$,
with little or no correlation between scan lines. Motivated by
these considerations and the extremely small height ranges in
the gels with 0$\%$ and $10\%$ surfactant, we constructed a
one-dimensional analysis procedure which analyzes the images one
scan line at a time, where a scan line is a line in the fast ($x$)
scan direction.

One-dimensional surface widths $\langle w_l \rangle$ were calculated in a
manner equivalent to that defined in Eq.~(\ref{eq:widthl}) with
$j$ now representing a sub-interval in the fast scan direction
instead of a sub-image and $(M/n)^2$ replaced by $(M/n)$.
The $w_{lj}$, for all $j$ and all scan lines, were then
averaged to get the $\langle w_l \rangle$ for the image.
The surface width calculations at the
smallest image sub-interval depends strongly on the definition of
$w_l$ (see Eq.~(\ref{eq:widthl})). 
When applied to a straight line ($\alpha = 1$), the
standard one-dimensional estimate of $\langle w_l \rangle$ 
used here overestimates $\alpha$ by as much as $20\%$ for 
sub-intervals that contain less than four data points. 
The results of the one-dimensional surface width analysis 
are presented in Table~\ref{tab:alpha} and in Sec.~\ref{sec:results}.

While surface-width calculations contain information about 
all length scales $\leq{L}$, calculation of correlation
functions yield information at a specific value of
relative separation. We define the height-height correlation
function of a scan line at fixed $y$ as
\begin{equation} C(r,y)=\langle h(x,y)h(x+r,y)
\rangle_{x} - \langle h(x,y) \rangle^{2}_x \;, \label{eq:ccorr}
\end{equation}
where $r$ is a relative displacement in the $x$ direction, and
$\langle \cdot \rangle_x$ denotes an average over all values of $x$ in
the scan line. Note that after flattening, the second term is
zero, and that $C(0,y)$ is the rms width of the scan line, $w(y)$.
Although the flattening was necessary to compensate for
non-horizontal mounting of the gel and/or gradients in the gel
thickness, it introduces correlations at the length scale $L$.
Consequently, correlation functions for flattened images 
are meaningful only for lengths, $l\leq{L/2}$. 
For scaling analysis, it is more convenient to deal
with an alternative form of the correlation function, which is
sometimes referred to as the variance of increments \cite{feder88}
or the increment correlation function,
\begin{equation}
H(r,y)=\langle |h(x+r,y)-h(x,y)|^q\rangle_x \;.
\label{eq:hcorr}
\end{equation}
For the special case of $q=2$,
\begin{equation}
H(r,y)=\langle [h(x+r,y)-h(x,y)]^2\rangle_x=2[w(y)^2-C(r,y)]\;.
\end{equation}
The correlation functions in the slow ($y$) scan direction,
$C(x,r)$ and $H(x,r)$, can be analogously defined. 

To improve the statistical accuracy of the increment correlation
function measurements, multiple scan lines must be used. Thus, we
define the first and second moments as
\begin{equation}
\begin{array}{l}
H_{\rm fast}(r)=\langle H(r,y) \rangle=\sum_{j=1}^p\left[\sum_{i=1}^{M} 
H(r,y=iL/M)\right]/{pM}\;,\\
\langle H(r,y)^2 \rangle=\sum_{j=1}^p\left[\sum_{i=1}^{M} 
H(r,y=iL/M)^2\right]/{pM}\;,
\end{array}
\label{eq:hmom}
\end{equation}
where $\sum_{j=1}^p$ runs over the $p$ different scan lines. 
The standard error of $\langle H(r,y)
\rangle$ is then $\sigma_{\rm fast}=\sigma(r,y)=\sqrt{\frac{\langle H(r,y)^2
\rangle-\langle H(r,y) \rangle^2}{p M-1}}$. Again, the moments
and error in the slow scan direction can be analogously defined.

\begin{figure}
{\epsfysize=3in \epsfbox{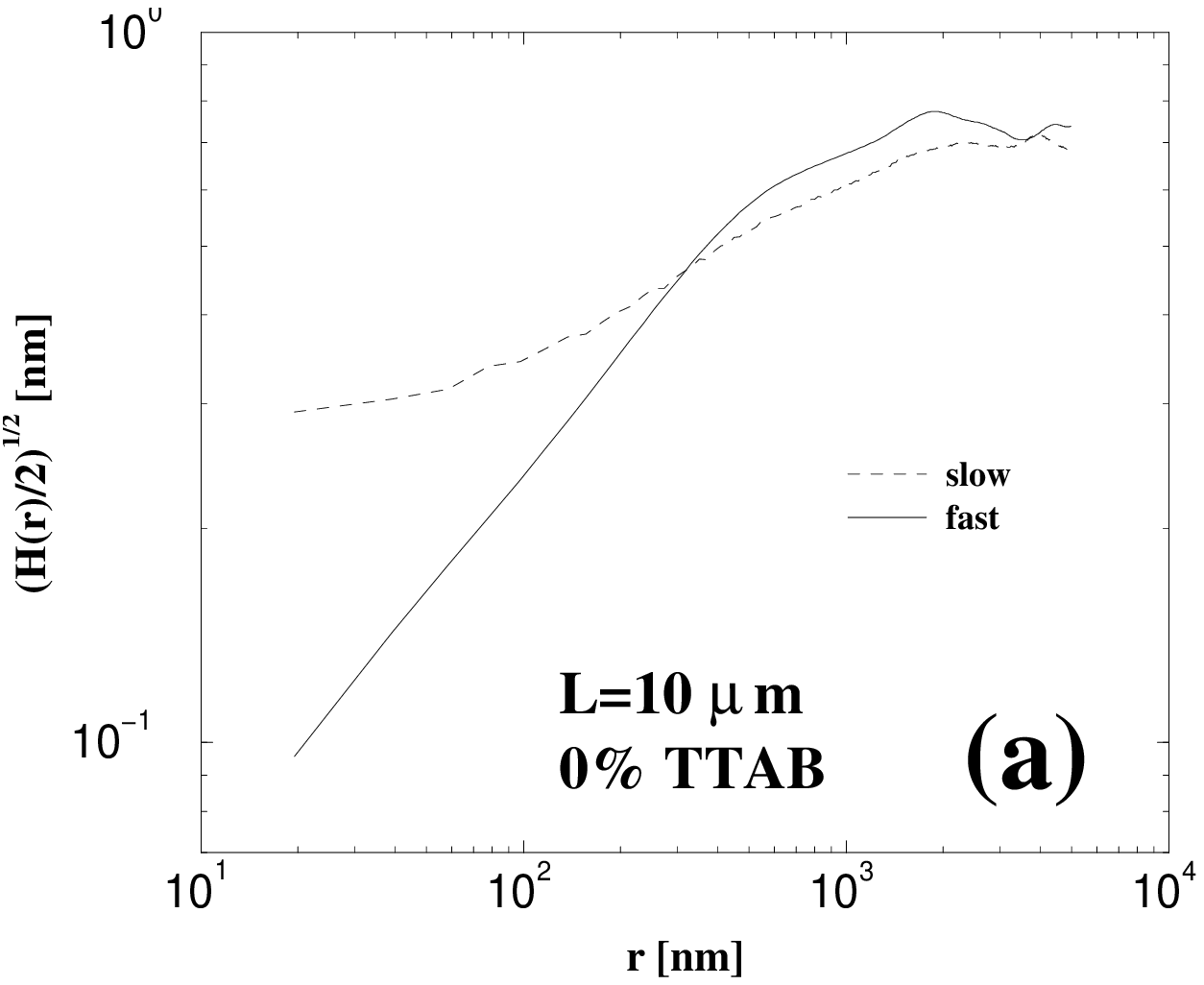}} 
{\epsfysize=3in \epsfbox{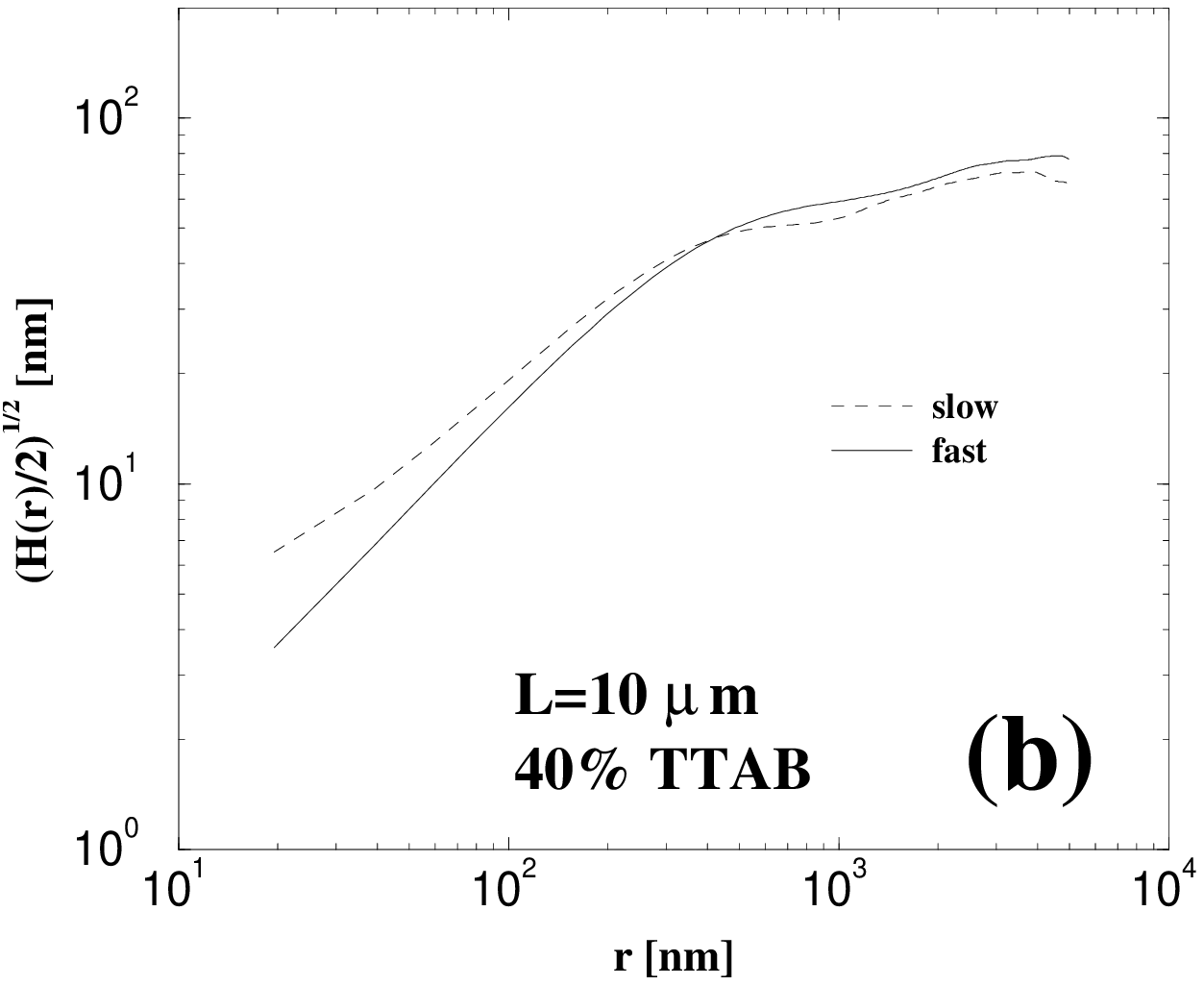}} 
\caption{ 
Plots of the one-dimensional $H_{\rm slow}(r)$
and $H_{\rm fast}(r)$ vs $r$ on a log-log scale for $L=10~\mu
\rm{m}$ (a) $0\%$ and (b) $40\%$ templated gel surfaces. It is
clear that for the untemplated gel surface the signal-to-noise
ratio is poor. For the $40\%$ templated gel surface, the
noise-induced random offsets between scan lines are smaller, but
not negligible.
} 
\label{fig:hslowhfast}
\end{figure}

The increment correlation function for $q=2$ is expected to scale as
\begin{equation}
H_{\rm fast}(r) \propto
\left\{
\begin{array}{ll}
r^{2 \alpha} & r \ll l_{\times} \\
2 w^{2}_{\rm sat} & r \gg l_{\times}
\end{array}
\right. \; .
\end{equation}
In Fig.~\ref{fig:1dhcorr} we see power-law scaling for small
$r$ and an apparent cross-over to saturated values at large $r$.
When the noise-induced offsets between scan lines are small,
$H_{\rm slow}(r)$ should have the same scaling as $H_{\rm
fast}(r)$. However, if the signal-to-noise ratio is small, we
expect $H_{\rm slow}(r)$ to measure only the noise with $\alpha = 1/2$.
Figure~\ref{fig:hslowhfast} shows plots of $H_{\rm slow}(r)$ and
$H_{\rm fast}(r)$ for $0\%$ and $40\%$ templated gel surfaces. For
the untemplated gel surfaces the fluctuations due to noise in the
slow scan direction are comparable to the signal. Thus $H_{\rm
slow}(r)$ does not follow $H_{\rm fast}(r)$. For the $40\%$
templated gel surfaces, the random offsets between scan lines are
relatively smaller, but not negligible. Hence $H_{\rm slow}(r)$ is
much closer to $H_{\rm fast}(r)$.

\begin{figure}
{\epsfxsize=4in \epsfbox{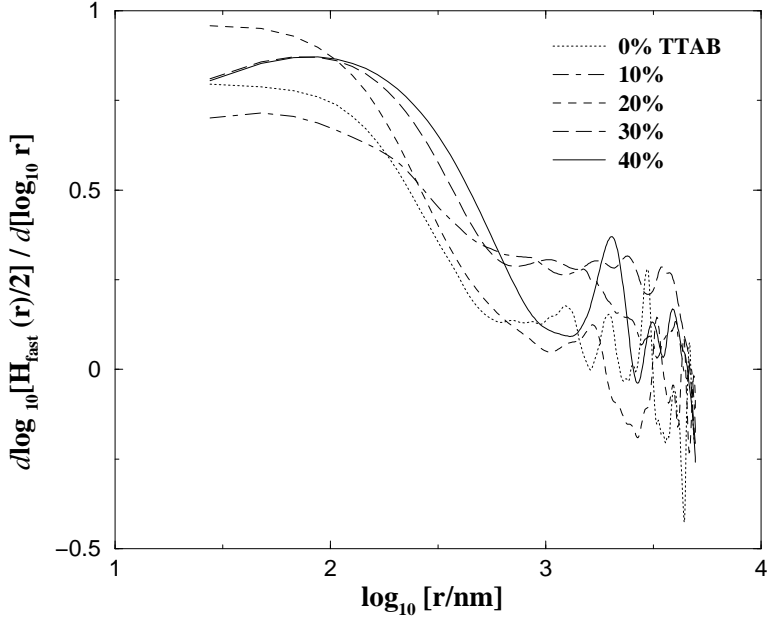}}
\caption{ 
Plot of $d \log_{10}{[H_{\rm fast}(r)/2]}/d [\log_{10} r]$ for $L = 10~\mu$m. 
The inflection point of this curve gives an estimate for the lateral
cross-over length, $l_{\times}$, which is listed in Table~\ref{tab:alpha}.
}
\label{fig:dhdr}
\end{figure}

By evaluating the increment correlation function and plotting
$H_{\rm fast}(r)$ on a log-log scale, the roughness exponent
$\alpha$, the limiting surface width $w_{\rm sat}$, and the
lateral cross-over length $l_{\times}$ can be determined from the
slope in the small-$r$ linear regime (see Fig.~\ref{fig:1dhcorr}),
the limiting value of $H_{\rm fast}(r)$ for large $r$, and the
inflection point of the logarithmic derivative of $H_{\rm fast}(r)$ (see
Fig.~\ref{fig:dhdr}), respectively.

\section{Results and Discussion}
\label{sec:results}

\begin{figure}
{\epsfysize=3in \epsfbox{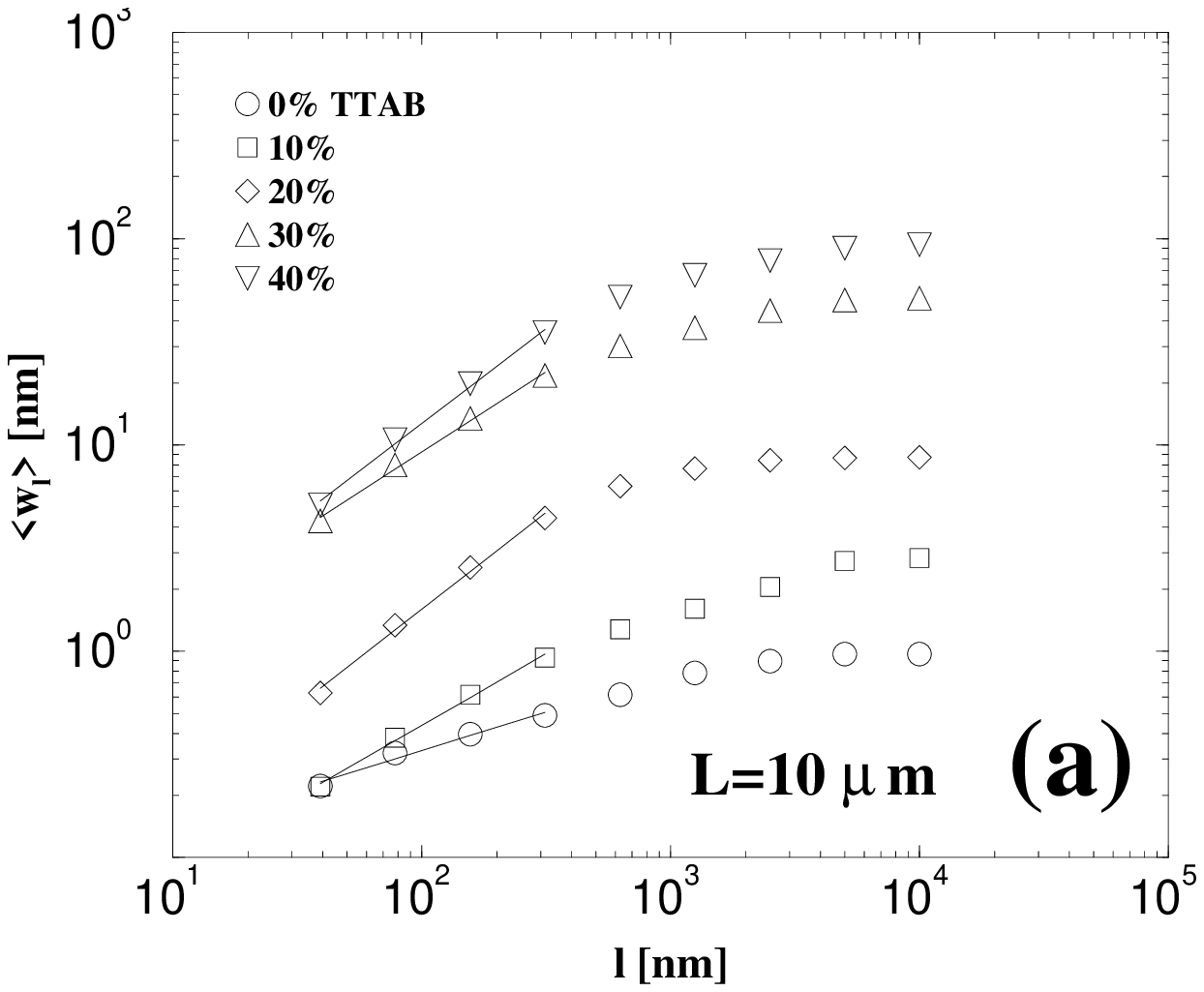}} 
{\epsfysize=3in \epsfbox{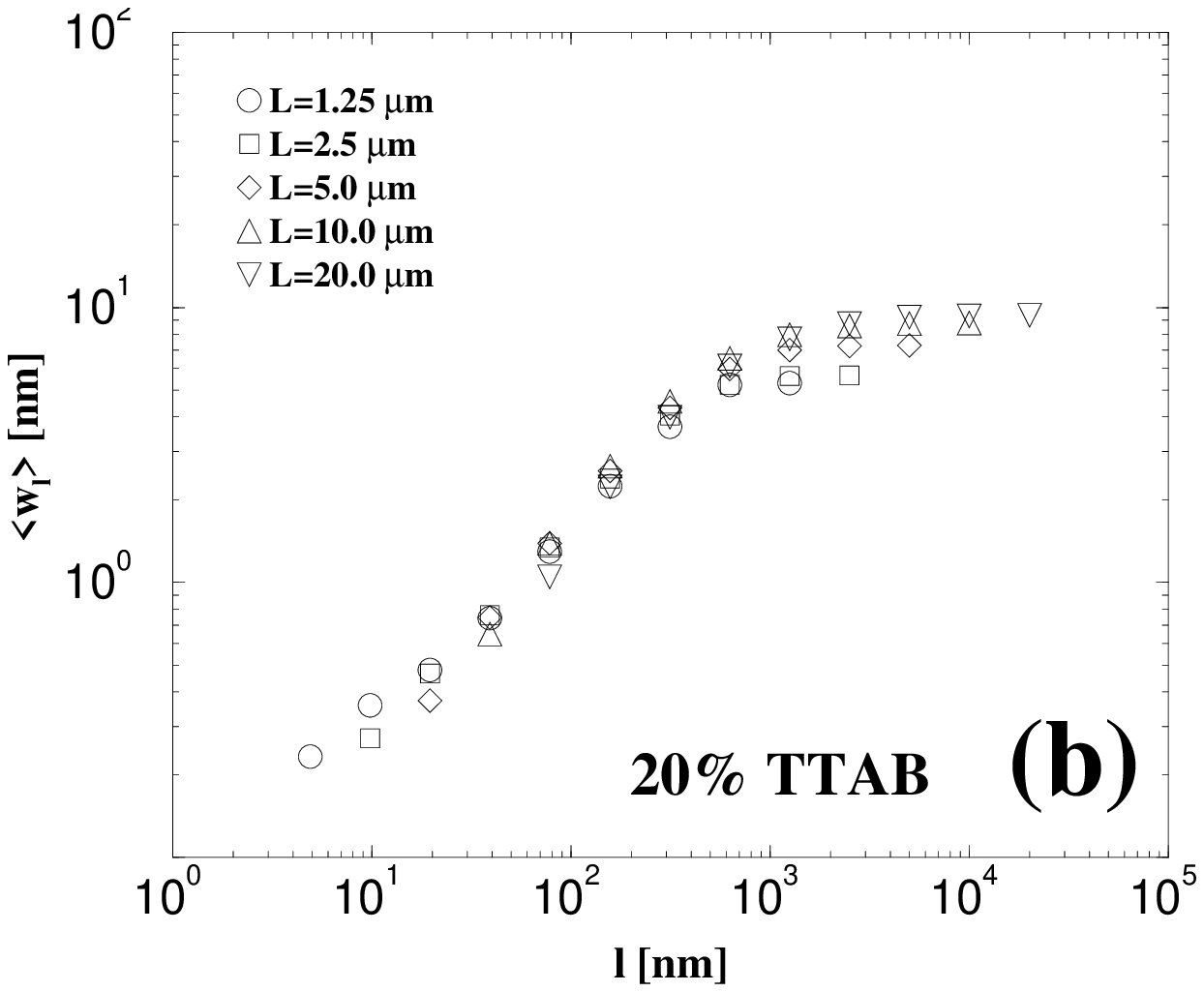}} 
\caption{ 
A typical plot of
$\langle w_l \rangle $ vs\ $l$ on a log-log scale, obtained from
two-dimensional images for (a) gel surfaces of various surfactant
concentrations (corresponding to Fig.~\protect\ref{fig:images10}) and (b)
$20\%$ templated gel surfaces at various scan sizes $L$
(corresponding to Fig.~\protect\ref{fig:zoom}). The solid lines in (a)
represent linear fits to selected data indicating power-law
behavior for limited ranges in $l$. Note the nice overlap of data
for various scan sizes in (b).
} 
\label{fig:2dwidth}
\end{figure}

\begin{figure}
{\epsfysize=3in \epsfbox{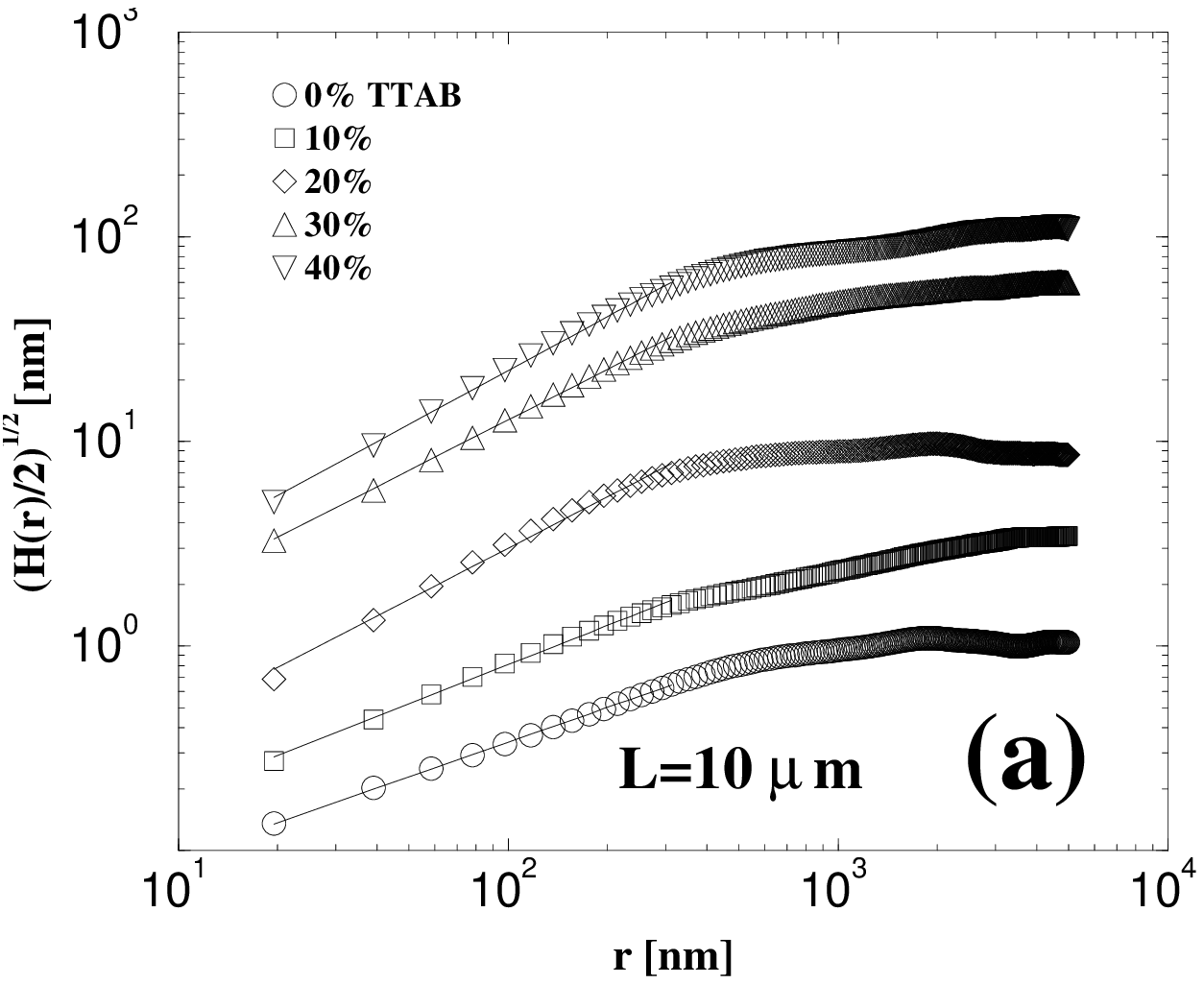}} 
{\epsfysize=3in \epsfbox{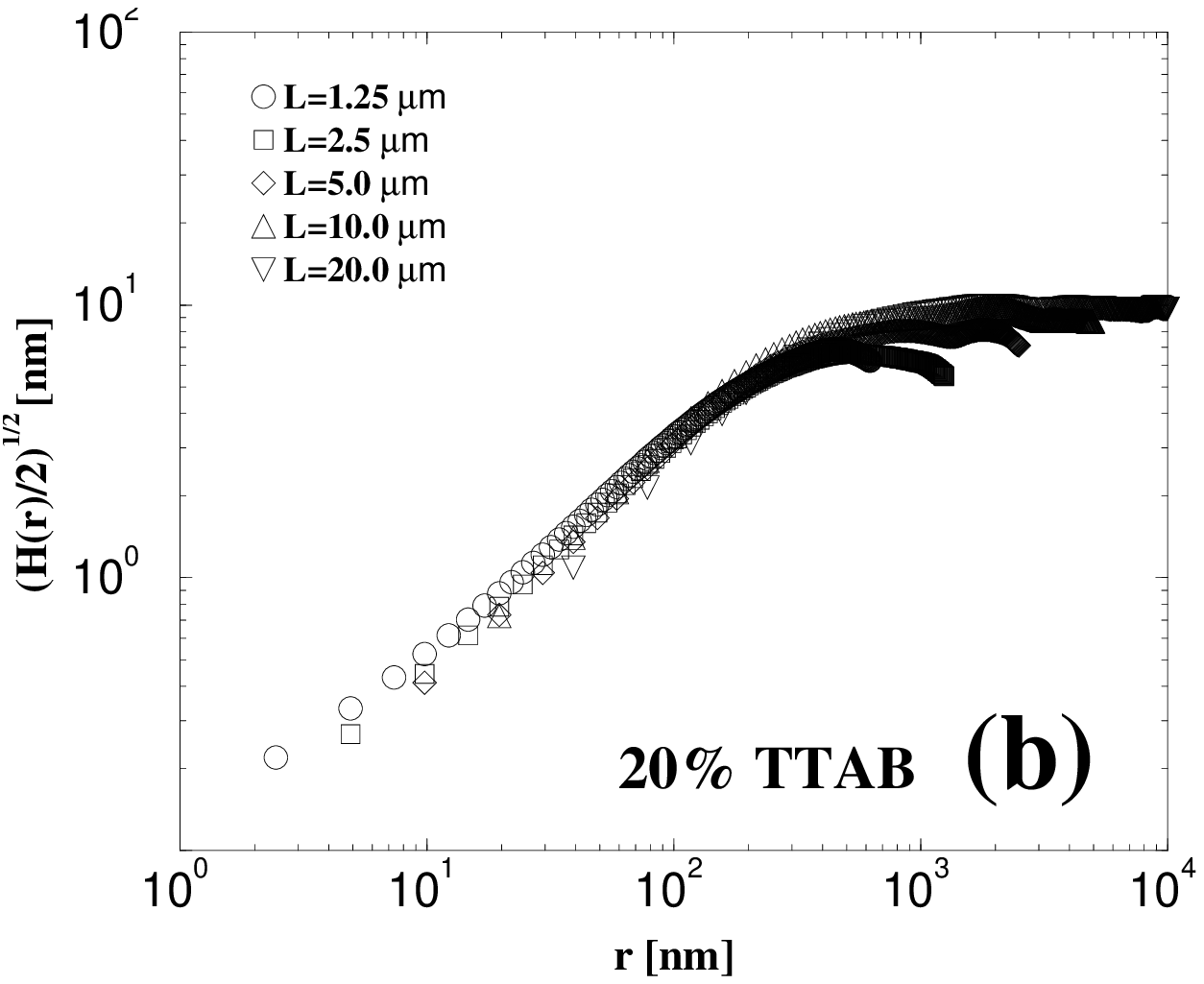}} 
\caption{ 
A typical plot of the one-dimensional
$H_{\rm fast}(r)$ vs $r$ on a log-log scale for (a) gel surfaces
of various surfactant concentrations at scan size $L = 10~\mu
\rm{m}$ (corresponding to Fig.~\protect\ref{fig:images10}) and (b) $20\%$
templated gel surfaces at various scan sizes $L$ (corresponding to
Fig.~\protect\ref{fig:zoom}). The solid lines in (a) represent linear fits
to selected data indicating power-law behavior for limited ranges
in $r$. Note the nice overlap of data for various scan sizes in (b). 
This figure should be compared with Fig.~\protect\ref{fig:2dwidth}. 
} 
\label{fig:1dhcorr}
\end{figure}

As shown in the previous sections, among the two-dimensional analysis
methods, only the $l$-dependent surface width $\langle w_l \rangle$ 
gives meaningful results for this system. The $l$-dependence of 
this quantity was presented in Fig.~\ref{fig:2dwidth} and is
essentially similar to that of the one-dimensional
surface-width (not shown) and the one-dimensional 
increment correlation function shown in Fig.~\ref{fig:1dhcorr}. 
All three quantities show a self-affine scaling region on short length
scales, and a saturated plateau on large length scales. These two
scaling regimes are separated by the crossover length $l_\times$. Both
the crossover length and the saturation value increase with increasing
surfactant concentration. 

\begin{figure}
{\epsfxsize=4in \epsfbox{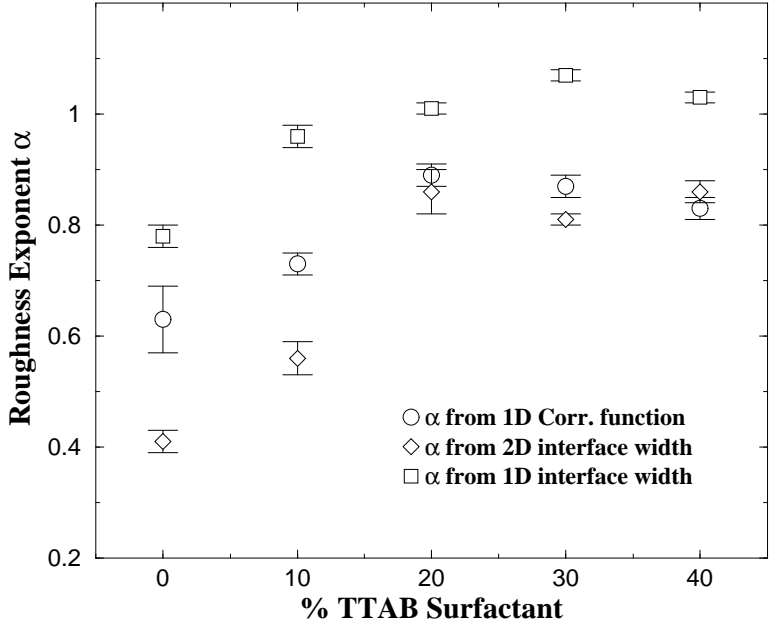}}
\caption{ 
Roughness exponent $\alpha$ vs surfactant
concentration. The measured values of $\alpha$ are averages over the
different scan sizes for each surfactant concentration.
The error bars represent the standard error. 
}
\label{fig:alpha}
\end{figure}

The values of the roughness exponent $\alpha$ extracted from the
self-affine regime for the 
surface width measurements (both two-dimensional and one-dimensional), 
along with those calculated from the one-dimensional
increment correlation function with $q=2$, are listed in Table~\ref{tab:alpha}. 
Using $q \neq 2$ in Eq.~(\ref{eq:hcorr})
yielded the same values of $\alpha$ as obtained with $q=2$. 
This indicates that the gel surfaces 
are perfectly self-affine on these short length scales, rather than
multi-affine \cite{meakin98,BUEN02,BARA91}.
A plot of the estimated values of $\alpha$ 
vs the surfactant concentration is shown in
Fig.~\ref{fig:alpha}. The estimates 
obtained from the two-dimensional surface width measurements and from the
one-dimensional increment correlation function agree well for $20\%$, $30\%$,
and $40\%$ templated surfaces. We believe that the lack of
agreement for $0\%$ and $10\%$ templated surfaces is associated with the
low signal-to-noise ratio in the images for these gels. The
vertical resolution limit of the AFM ($\agt 0.1~\rm{nm}$
\cite{crmdi3000}) contributes to the uncertainty since the height
variations are only about 1~nm for the lowest surfactant concentrations.

The values for $\alpha$, determined from the one-dimensional surface-width
measurements, are consistently higher than those obtained from the
other two measurements. As discussed in Sec.~\ref{sec:1d}, the
roughness exponents obtained using the one-dimensional surface width 
for intervals with only a few data points are consistently too high, and
this may be at least part of the explanation for the discrepancy. 
The values of $\alpha$
for $20\%$, $30\%$, and $40\%$ templated surfaces obtained
from two-dimensional surface-width and one-dimensional 
correlation-function measurements  are clustered between 0.80 and 0.89. 
However, measurements of surface-roughness exponents from relatively
small data sets, such as the present ones, tend to have systematic errors,
as well as random ones, and the calibration curves for synthetic data
sampled over 512 points, given in Ref.~\cite{CAST01}, indicate that
the values obtained here may be underestimated by up to $10\%$. 
Thus, based on our results it is not unreasonable to conclude that all
of these surfaces have the same roughness exponent in the range 0.85 to
1.0, irrespective
of the surfactant concentration. 
A roughness exponent in this range is consistent with previous results 
for N-isopropylacrylamide gel surfaces by Suzuki 
{\em et al.}~\cite{suzuki97}.
However, as discussed in the next
paragraph, the {\it magnitude} of the height variations depends strongly
on the surfactant concentration, as do the crossover length $l_\times$ 
and the saturation surface width $w_{\rm sat}$.

\begin{figure}
{\epsfxsize=4in \epsfbox{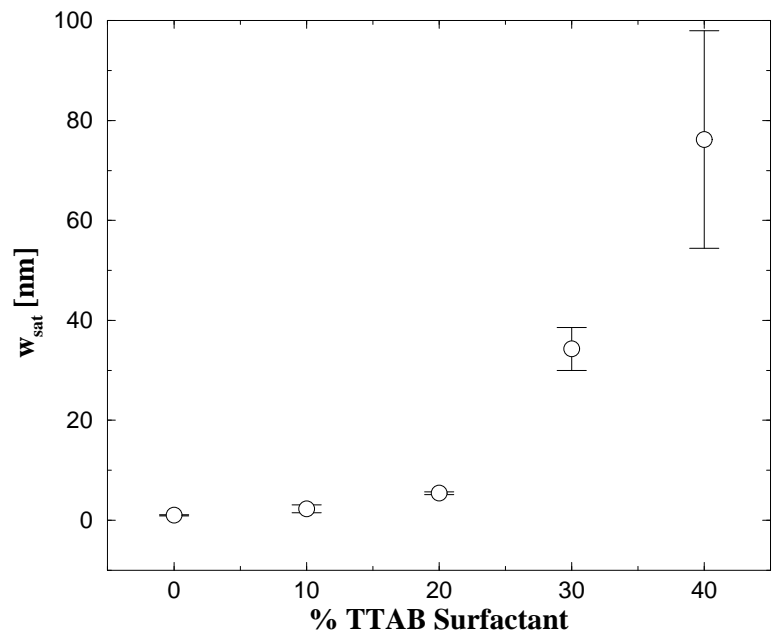}}
\caption{ 
Saturation surface width $w_{\rm sat}$ vs surfactant concentration.
The error bars represent the standard error.
}
\label{fig:wsat}
\end{figure}

\begin{figure}
{\epsfxsize=4in \epsfbox{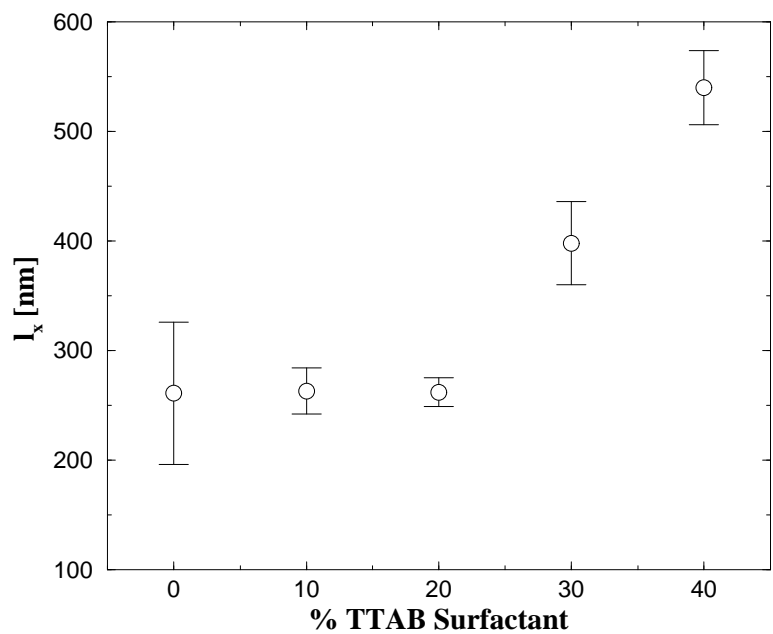}}
\caption{ 
Cross-over length $l_{\times}$ vs surfactant concentration.
The error bars represent the standard error.
}
\label{fig:lcross}
\end{figure}

The limiting value of the surface width on large length scales, $w_{sat}$, 
is found to increase by almost two orders of magnitude as the surfactant
concentration is increased from 0\% to 40\%. This behavior is evident from
Figs.~\ref{fig:scanline}, \ref{fig:2dwidth}(a),
\ref{fig:1dhcorr}(a), and~\ref{fig:wsat}. The measurements of
$w_{sat}$ are listed in Table~\ref{tab:alpha} and displayed vs surfactant 
concentration in  Fig.~\ref{fig:wsat}. 
Similarly, but less spectacularly, $l_\times$
increases by about a factor of two, as shown in Table~\ref{tab:alpha} 
and Fig.~\ref{fig:lcross}. 

As the surfactants are removed, the transparency of some of the gels
change. The gels that contained no or little surfactant stay
clear. On the other hand, gels that had higher concentrations of
surfactant (more than $20\%$) become opaque and white as the
surfactant diffuses out. 
Using the theory of light scattering by refractive index fluctuations, 
an expression for the spectral density of 
light scattered can be obtained. 
Reasonable functional forms for the correlation function can then be 
used to obtain the turbidity or inverse attenuation length.
A detailed derivation is provided in the Appendix.  
The turbidity for a wavelength $\lambda=550$~nm in the visible 
region is shown as a function of the correlation length~$\xi$ in 
Fig.~\ref{fig:scattint}. 
The two curves presented correspond to an isotropic exponential and 
Gaussian correlation function, respectively.
If we associate the correlation length $\xi$ with the saturation 
value of the rms surface width, $w_{\rm sat}$, 
we notice that the turbidity increases by about six orders of 
magnitude as $w_{\rm sat}$ changes from 1 nm for the $0\%$ 
templated gel to 80 nm for the $40\%$ templated gel.
These changes in the turbidity are consistent with a change from a 
transparent to an opaque material. 
Our justification for the association between $\xi$ and $w_{\rm sat}$ 
is discussed below.

The experimental results presented here indicate that while the templated 
gel surfaces are self-affine on small to moderate length scales, with 
$\alpha$ in the range 0.85 to 1, both the crossover length and the saturation 
value of the surface width increase significantly with the bulk concentration 
of surfactant. From Figs.~\ref{fig:wsat} and~\ref{fig:lcross} 
it is clear that most of these increases occur only after the surfactant 
concentration exceeds about  
20\%. This indicates that the dramatic changes seen in the surface 
structure are reflections of a sharp transition in the bulk structure, most 
likely a percolation transition at which the surfactant structure changes from 
a spatial distribution of individual micelles to a percolating network. These 
different structures apparently do 
not seriously affect the mechanical properties of the 
gel as long as the surfactants are in place. However, once they are soaked out, 
the resulting percolating void network is likely to weaken the gel and lead to 
large relaxations as the elastic polymer network readjusts its configuration. 
While the percolation threshold is a non-universal quantity and thus 
depends on the particular system, 
for site percolation on standard three-dimensional lattices it lies roughly 
in the range from 20 to 30 percent by volume~\cite{STAU92}. This is 
consistent with the range in which the surface structure and the opacity of 
the templated gels undergo their strongest changes. 
Previous work on the phase diagrams of alkyltrimethylammonium 
surfactants~\cite{jonsson88}
suggests a transition from micelle to hexagonal phase at around 
$35~\rm{wt}\%$ TTAB at room temperature.

Further support for the percolation hypothesis is provided by recent computer 
simulations of a two-dimensional frustrated spring network with a free 
surface and a variable density of holes \cite{BUEN01,BUEN02}, 
which were inspired by the experimental results presented here. In those 
simulations the structure of the free surface remains relatively constant, 
with a self-affine structure on short length scales and a finite saturation 
width on large length scales, as long as 
the volume fraction of holes is far below the percolation limit. 
As the hole concentration approaches the percolation threshold, both the 
the saturation width and the crossover length increase in a manner qualitatively 
similar to our experimental observations. 
The saturation width, in particular, appears to follow the same power-law 
divergence as the bulk correlation length over a wide range of hole 
concentrations approaching the percolation threshold.
These observations form our justification for the association of $w_{\rm sat}$ 
with the bulk correlation length in the estimate of the change in the 
turbidity, discussed above.

\section{Conclusion}
\label{sec:conc}

We have obtained AFM images of relaxed, free surfaces
of PAAm hydrogels templated with TTAB surfactant and 
performed a numerical scaling analysis of the observed surface structure. 
In doing so, we have developed methods that can be used not only for our data,
but also to perform scaling analysis of any surface imaged by an
AFM. We showed that the calculation of the two-dimensional
surface width and one-dimensional correlation functions gave
consistent, useful results. Two-dimensional box-counting and
spectral methods were unsatisfactory, while results from
one-dimensional surface-width measurements provided
over-estimates at small length scales. The one-dimensional
increment correlation function in the fast scan direction, 
$H_{\rm fast}(r)$, is simple to
calculate and gives a reliable characterization of rough surfaces.
By calculating this correlation function we have determined that
the surfaces of the templated gels are self-affine over a
significant range of small and intermediate length scales, with a roughness 
(Hurst) exponent in the range 0.85 to 1. This self-affine regime is terminated 
by a cross-over length, above which the surface width is scale independent. The
increases in the cross-over length and the saturation width of the surface 
with increasing surfactant concentration  
coincides with the observation of a large increase in the scattering of visible 
light. We propose that the observed 
effects are caused by a percolation transition
in the bulk surfactant system, from isolated micelles to a percolating 
network. Further investigations of the bulk structure of the gels at different 
surfactant concentrations are planned. 

\section*{Acknowledgments}

We thank R.~L.\ Rill, G.~M.\ Buend{\'\i}a, P.~Meakin, and S.~Hong for useful
discussions, and E.~Lochner for his assistance with the AFM and the
development of the new sample holder. 
G.~Brown, K.~Park, and S.~Wang are acknowledged for their comments on the manuscript.
We thank B.~van de Burgt for the Raman spectroscopy results and analysis, and
A.~Beheshti is acknowledged for his help in testing the validity of our data-analysis code.
Supported in part by NSF grants No.\ BES-951381 and DMR-9981815, and by Florida
State University through the Center for Materials Research and
Technology and the School of Computational Science and Information Technology.

\appendix
\section{}
The theory of light scattering by refractive-index fluctuations leads, 
in the Born approximation, to
the following expression for the spectral density of light scattered 
from a volume $V$,
a distance $R$ away in the far-field zone~\cite{berne00}:
\begin{equation}
I(\vec{k})= I_{0}\frac{V}{R^2} \frac{(2\pi)^4 \sin^2 \psi}{2\lambda^4} 
{\cal S}^{(\delta n)}(\vec{k}) \;,
\label{eq:specintensity}
\end{equation}
where
\begin{equation}
{\cal S}^{(\delta n)}(\vec{k}) = \int_{V}d^{3}r\, G(\vec{r})\exp[i
\vec{k}\cdot\vec{r}]\;.
\end{equation}
Here $I_{0}$ is the intensity of the incident light, 
$\lambda$ is the wavelength of the scattered light, $\psi$ 
is the polarization angle,
$\vec{k}$ is the scattering wave vector, and $G(\vec{r})=
\langle\delta n(\vec{r}_0)\delta n(\vec{r}_0 + 
\vec{r})\rangle/\langle(\delta n)^2\rangle$ is
the normalized correlation function for the refractive-index 
fluctuations $\delta n = n-\langle n \rangle$. 

\begin{figure}
{\epsfxsize=4in \epsfbox{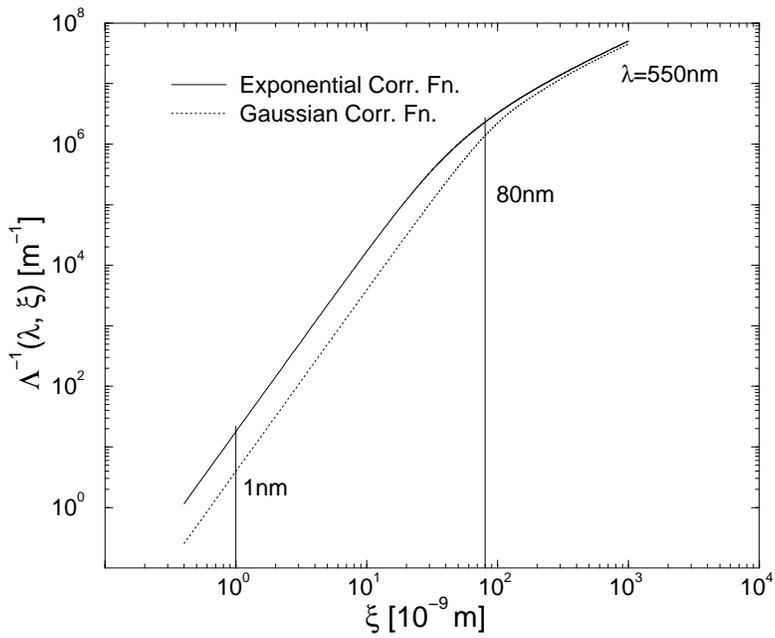}} 
\caption{ 
Turbidity ${\Lambda^{-1}}(\lambda, \xi)$ vs correlation length $\xi$ for
wavelength $\lambda = 550$~nm in the visible region. 
The solid and dotted curves represent an exponential and a Gaussian form for the
correlation function, respectively. 
Note that the turbidity changes by about six orders of magnitude
as the correlation length changes from $1$~nm 
(corresponding to $w_{\rm sat}$ for a $0\%$ templated gel) to $80$~nm 
(corresponding to $w_{\rm sat}$ for a $40\%$ templated gel). 
This increase in turbidity indicates a change from a transparent to an opaque material.} 
\label{fig:scattint}
\end{figure}

We can calculate the total scattering intensity at a particular
wavelength $\lambda$ by using the relation $|\vec{k}|
=\frac{4\pi}{\lambda}\sin(\frac{\theta}{2})$, where $\theta$ is 
the scattering angle, and integrating $I(\vec{k})$ over all
angles, \textit{i.e.},
\begin{equation}
I(\lambda)= I_{0}\frac{V}{R^2} \Lambda^{-1}(\lambda, \xi)
\end{equation}
where the turbidity or inverse attenuation length is
\begin{equation}
\Lambda^{-1}(\lambda, \xi) = \int d\theta d\phi\sin\theta\,
\frac{1}{2}(1+\cos^2\theta)\, {\cal S}^{(\delta n)}(\vec{k})\;,
\label{eq:scattintensity}
\end{equation}
and $\xi$ is the bulk correlation length. The term in the integral of the form 
$(1+\cos^2\theta)/2$ corresponds to unpolarized incident light \cite{johnson94}.
Assuming an isotropic exponential correlation function, $G(r) =
\exp{[-r/\xi]}$, we find
\begin{equation}
\Lambda^{-1}(\lambda, \xi)= \frac{4\pi^{3}\langle(\delta n)^2\rangle}
{\lambda}\frac{a^2+2}{a}\left[ \frac{a^2+2}{a^2+1}
- \frac{2~\ln{(a^2+1)}}{a^2} \right]\;,
\label{eq:expcorr}
\end{equation}
where $a=4\pi \xi/\lambda$.
Alternatively, using a Gaussian correlation function, 
$G(r) = \exp{[-(r/\xi)^2]}$, we find
\begin{equation}
\Lambda^{-1}(\lambda, \xi) = \frac{2\pi^{7/2}\langle(\delta n)^2\rangle }{\lambda a^3}
\left\{(64-8a^2+a^4) - (64+8a^2+a^4) \exp{[-a^2/4]}\right\}\;.
\label{eq:gausscorr}
\end{equation}
A reasonable value for the mean refractive-index fluctuation between 
the PAAm and water regions in a gel is 
$\langle (\delta n)^2\rangle = 0.01$. 
The resulting turbidity, as given by Eqs.~(\ref{eq:expcorr}) and~(\ref{eq:gausscorr}), 
is shown in Fig.~\ref{fig:scattint}.

\clearpage
\nonumber \noindent

\begin{table}
\caption{Number of data sets (images) used for analysis.}
\begin{tabular}{|r|r|r|r|r|r|r|}
\hline
& \multicolumn{6}{l|}{Scan Size $L$~[$\mu$m]} \\
\hline
\% Surfactant & 1.25 & 2.5 & 5.0 & 10.0 & 20.0 & Total\\
\hline
0 \% &  6 & 5 & 4 & 3 & 2 & 20 \\
\hline
10 \% & 5 & 5 & 5 & 3 & 5 & 23 \\
\hline
20 \% & 5 & 3 & 5 & 5 & 6 & 24 \\
\hline
30 \% & 4 & 5 & 5 & 4 & 2 & 20 \\
\hline
40 \% & 4 & 5 & 4 & 4 & 2 & 19 \\
\hline
\multicolumn{6}{|r|}{Grand Total:} & 106\\
\hline
\end{tabular}
\label{tab:sets}
\end{table}

\begin{table}
\caption{Summary of results of scaling analyis for templated gel
surfaces.}
\begin{tabular}{|c|c|c|c|c|c|}
\hline
\% Surfactant & \multicolumn{3}{l|}{Roughness Exponent $\alpha$, from:} 
& RMS Width & Cross-over length \\
\cline{2-4}
&Two-dimensional $\langle w_l \rangle$ &One-dimensional $\langle w_l \rangle$ & $H_{\rm fast}(r)$ 
&$w_{sat}$ [nm] & $l_{\times}$ [nm] \\
\hline
0 \% &  0.41$\pm$0.02 & 0.77$\pm$0.02 
& 0.63$\pm$0.06 & 1.3$\pm$0.1 & 260$\pm$70 \\
\hline
10 \% & 0.56$\pm$0.03 & 0.90$\pm$0.02 
& 0.73$\pm$0.02 & 2.3$\pm$0.4 & 260$\pm$20 \\
\hline
20 \% & 0.86$\pm$0.02 & 1.009$\pm$0.006 
& 0.89$\pm$0.02 & 7.4$\pm$0.3 & 260$\pm$10 \\
\hline
30 \% & 0.80$\pm$0.01 & 1.023$\pm$0.005 
& 0.87$\pm$0.02 & 41$\pm$3 & 400$\pm$40 \\
\hline
40 \% & 0.86$\pm$0.02 & 1.026$\pm$0.005 
& 0.83$\pm$0.02 & 80$\pm$10 & 540$\pm$30\\
\hline
\end{tabular}
\label{tab:alpha}
\end{table}

\clearpage
\nonumber \noindent


\end{document}